# A General Systems Theory for Chaos, Quantum Mechanics and Gravity for Dynamical Systems of all Space-Time Scales

A M Selvam
Deputy Director (Retired)
Indian Institute of Tropical Meteorology, Pune 411 008, India
Email: amselvam@eth.net Web site: http://www.geocities.com/amselvam

**Abstract**
Non-local connections, i. e. long-range space-time correlations intrinsic to the observed subatomic dynamics of quantum systems is also exhibited by macro-scale dynamical systems as selfsimilar fractal space-time fluctuations and is identified as self-organized criticality. The author has developed a general systems theory for the observed self-organized criticality applicable to dynamical systems of all space-time scales based on the concept that spatial integration of enclosed small-scale fluctuations results in the formation of large eddy circulation. The eddy energy spectrum therefore represents the statistical normal distribution according to the Central Limit Theorem. The additive amplitudes of eddies, when squared (variance or eddy kinetic energy), represent the statistical normal (probability) distribution, a result observed in the subatomic dynamics of quantum systems. The model predicts Kepler's laws of planetary motion for eddy circulation dynamics. Inverse square law of gravitation therefore applies to the eddy continuum ranging from subatomic to macro-scale dynamical systems, e.g. weather systems. The model is similar to a superstring model for subatomic dynamics which unifies quantum mechanical and classical concepts and manifestation of matter is visualised as vibrational modes in string-like energy flow patterns. The cumulative sum of centripetal forces in a hierarchy of vortex circulations may result in the observed inverse square law form for gravitational attraction between inertial masses of the eddies.

PACS Category:   80.

## 1. Introduction

Atmospheric flows, a representative example of turbulent fluid flows, exhibit long-range spatiotemporal correlations manifested as the fractal geometry to the global cloud cover pattern concomitant with inverse power law form for spectra of temporal fluctuations. Such non-local connections are ubiquitous to dynamical systems in nature and are identified as signatures of self-organized criticality [1]. A review of research work on the applications of selfsimilarity and self-organized criticality in atmospheric sciences is given in [2]. A cell dynamical system model developed for atmospheric flows shows that the observed long-range spatiotemporal correlations are intrinsic to quantumlike mechanics governing fluid flows [3 - 5]. The model concepts are independent of the exact details such as the chemical, physical, physiological and other properties of the dynamical system and therefore provide a general systems theory applicable to all real world and computed dynamical systems in nature [6 - 13].

The model is based on the concept that spatial integration of enclosed small-scale fluctuations results in the formation of large eddy circulations. The model predicts the following: (i) the flow structure consists of an overall logarithmic spiral trajectory with the quasiperiodic *Penrose* tiling pattern for the internal structure (ii) conventional power spectrum analysis will resolve such spiral trajectory as a continuum of eddies with progressive increase in phase (iii) increments in phase angle are concomitant with increase in period length and also represents the variance, a characteristic of quantum systems identified as '*Berry's phase*' (iv) the universal algorithm for self-organized criticality is expressed in terms of the universal *Feigenbaum*'s constants [14] $a$ and $d$ as $2a^2=\pi d$ where, the fractional volume intermittency of occurrence $\pi d$ contributes to the total variance $2a^2$ of fractal structures (v) the *Feigenbaum*'s constants are expressed as functions of the *golden mean* (vi) the quantum mechanical constants '*fine structure constant*' and '*ratio of proton mass to electron mass*' which are pure numbers and obtained by experimental observations only, are now derived in terms of the *Feigenbaum*'s constant *a* (vii) atmospheric flow structure follows *Kepler*'s third law of planetary motion. Therefore *Newton*'s *inverse square law for gravitation* applies to eddy masses also. The centripetal accelerations representing inertial masses (of eddies) are equivalent to gravitational masses. Fractal structure to the spacetime continuum can be visualized as a nested continuum of vortex (eddy) circulations whose inertial masses obey Newton's inverse square law of gravitation. The model concepts are equivalent to a superstring model for subatomic dynamics, which incorporates gravitational forces. El Naschie [15] and Argyris and Ciubotariu [16] have discussed the fractal structure to spacetime and also state that fractalisation of micro-space is the origin of gravity.

## 2. A general systems theory for fluid flows: A string theory for fractal spacetime

The fractal spacetime fluctuations of dynamical systems may be visualized to result from the superimposition of an ensemble of eddies, namely an eddy continuum. The relationship between large and small eddy circulation parameters are obtained on the basis of Townsend's [17] concept that large eddies are envelopes enclosing turbulent eddy (small-scale) fluctuations (figure 1).

Figure 1. Physical concept of eddy growth process by the self-sustaining process of ordered energy feedback between the larger and smaller scales, the smaller scales forming the internal circulations of larger scales. The figure shows a uniform distribution of dominant turbulent scale eddies of length scale *2r*. Larger-eddy circulations such as ABCD form as coherent structures sustained by the enclosed turbulent eddies.

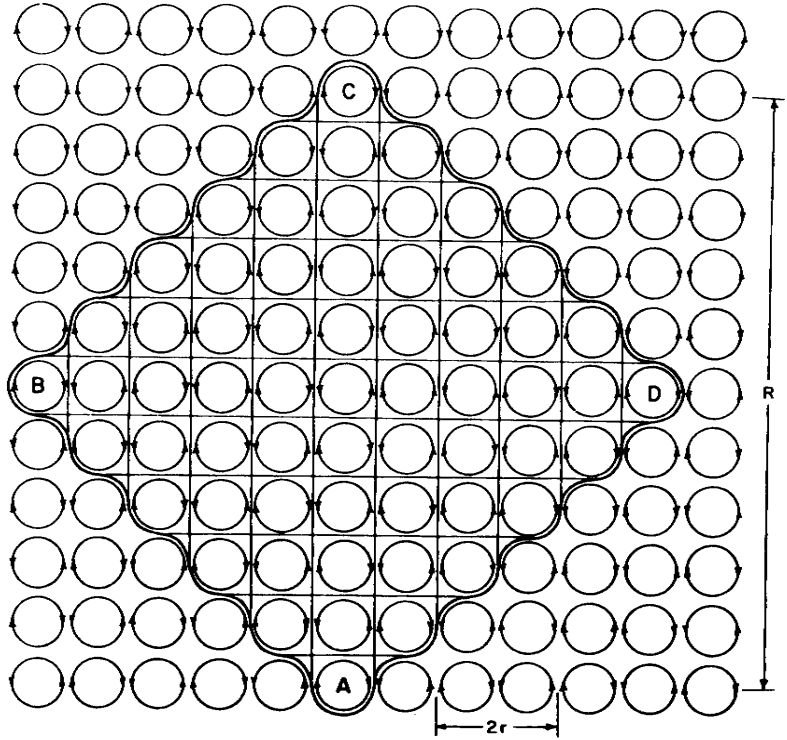

The relationship between root mean square (r. m. s.) circulation speeds $W$ and $w_*$ respectively of large and turbulent eddies of respective radii $R$ and $r$ is then given as

$$W^2 = \frac{2}{\pi}\frac{r}{R}w_*^2 \qquad (1)$$

The dynamical evolution of spacetime fractal structures is quantified in terms of ordered energy flow between fluctuations of all scales in (1), because the square of the eddy circulation speed represents the eddy energy (kinetic). A hierarchical continuum of eddies is generated by the integration of successively larger enclosed turbulent eddy circulations. The eddy energy (kinetic) spectrum then follows statistical normal distribution according to the *Central Limit Theorem* [18]. Therefore, square of the eddy amplitude or the variance represents the probability. Such a result that the additive amplitudes of eddies, when squared, represent the probability densities is observed for the subatomic dynamics of quantum systems such as the electron or photon [19]. Townsend's visualization of large eddy structure as quantified in (1) leads to the important result that the self-similar *fractal* fluctuations of atmospheric flows are manifestations of quantumlike chaos. Incidentally, one of the strangest things about physics is that we seem to need two different kinds of mechanics, quantum mechanics for microscopic dynamics of quantum systems and classical mechanics for macro-scale phenomena [20 - 23]. The above visualization of the unified network of atmospheric flows as a quantum system is consistent with Grossing's [21] concept of quantum systems *as order out of chaos* phenomena. Order and chaos have been reported in strong fields in quantum systems [24].

The square of the eddy amplitude $W^2$ represents the kinetic energy $E$ given as (1)

$$E = H\nu \qquad (2)$$

where $\nu$ (proportional to $1/R$) is the frequency of the large eddy and $H$ is a constant equal to $\frac{2}{\pi} r w_*^2$ for growth of large eddies sustained by constant energy input proportional to $w_*^2$ from fixed primary small scale eddy fluctuations. Energy content of eddies is therefore similar to quantum systems which can possess only discrete quanta or packets of energy content $h\nu$ where $h$ is a universal constant of nature (Planck's constant) and $\nu$ is the frequency in cycles per second of the electromagnetic radiation. The relative phase angle between large and turbulent eddies is equal to $r/R$ and is directly proportional to $W^2$ (1). The phase angle therefore represents variance and also there is progressive increase in phase with increase in wavelength. The above relationship between phase angle, variance and frequency has been identified as *Berry's Phase* [25] in the subatomic dynamics of quantum systems. Berry's phase has been identified in atmospheric flows [4].

Writing (1) in terms of the periodicities $T$ and $t$ of large and small eddies respectively, where

$$T = \frac{2\pi R}{W}$$

and

$$t = \frac{2\pi r}{w_*}$$

we obtain

$$\frac{R^3}{T^2} = \frac{2}{\pi} \frac{r^3}{t^2} \qquad (3)$$

The above equation (3) is analogous to *Kepler*'s third law of planetary motion, namely, the square of the planet's year (period) to the cube of the planet's mean distance from the *Sun* is the same for all planets [26, 27]. Newton developed the idea of an inverse square law for gravitation in order to explain Kepler's laws, in particular, the third law. Kepler's laws were formulated on the basis of observational data and therefore are of empirical nature. A basic physical theory for the inverse square law of gravitation applicable to all objects, from macro-scale astronomical objects to microscopic scale quantum systems is still lacking. The model concepts are analogous to a string theory [28] where, superposition of different modes of vibration in stringlike energy flow patterns result in material phenomena with intrinsic quantumlike mechanical laws which incorporate inverse square law for inertial forces, the equivalent of gravitational forces, on all scales of eddy fluctuations from macro- to microscopic scales. The cumulative sum of centripetal forces in a hierarchy of vortex circulations may result in the observed inverse square law form for gravitational attraction between inertial masses (of the eddies). Uzer *et. al* [29] have discussed new developments within the last two decades which have spurred a remarkable revival of interest in the application of classical mechanical laws to quantum systems. The atom was originally visualized as a miniature solar system based on the assumption that the laws of classical mechanics apply equally to electrons and planets. However within a short interval of time the new quantum mechanics of *Schrodinger* and *Heisenberg* became established (from the late 1920s) and the analogy between the structure of the atom and that of the solar system seemed invalid and classical mechanics became the domain of the astronomers. There is now a revival of interest in classical and semi-classical methods, which are found to be unrivaled in providing an intuitive and computationally tractable approach to the study of atomic, molecular and nuclear dynamics.

The apparent paradox of wave-particle duality in microscopic scale quantum systems [20] is however physically consistent in the context of macro-scale atmospheric flows since the bi-directional energy flow structure of a complete atmospheric eddy results in the formation of clouds in updraft regions and dissipation of clouds in downdraft regions. The commonplace occurrence of clouds in a row is a manifestation of *wave-particle duality* in the macro-scale quantum system of atmospheric flows (figure 2). Recent experiments by S. S. Afshar [30] have shown that the subatomic system, photon, exhibits both wave and particle characteristics at the same time. Recent theoretical studies indicate that the particle's mass can be thought of as extending throughout space-time as a wave, with the result that the global geometry of space - its curvature throughout the Universe depends on the properties of that wave [31].

The above-described analogy of quantumlike mechanics for atmospheric flows is similar to the concept of a subquantum level of fluctuations whose spacetime organization gives rise to the observed manifestation of subatomic phenomena, i. e. quantum systems as order out of chaos phenomena.

Puthoff [32] has shown that the observed stability of ground-state electronic orbits in atoms is a result of energy exchange with the sea of electromagnetic energy available in the vacuum zero point fluctuations. Quantum theory predicts that vacuum is constantly fizzing with particles that pop in and out of existence [33]. Historically, quantum mechanics had imposed arbitrary stability criterion for the ground state of electron orbits. Stable ground state is not possible in classical physics since attractive forces between the negative electron and positive nucleus will result in spiraling of orbital electrons into the nucleus accompanied by loss of energy due to emission of radiation by the accelerating electron, since all accelerating charges radiate energy. Puthoff [34] has also put forth the concept of "gravity as a zero-point fluctuation force". The vacuum zero-point fluctuation (electromagnetic) energy is manifested in the *Casimir effect* [35], namely a force between two closely spaced metal plates. Casimir effect is interpreted as due to imbalances in the zero-point energy caused by the presence of the plates and is analogous to the turbulent scale fluctuations whose spatial integration results in coherent large eddy structures. Recent studies show that background noise enhances weak signals in electronic circuits [36]. El Naschie [15] has proposed in a series of papers that *Cantorian-fractal* conception of spacetime may effect reconciliation between quantum mechanics and gravity. Castro and Granik [37] propose that the world emerged as a result of a nonequilibrium process of self-organized critical phenomena launched by vacuum fluctuations in Cantorian–fractal space-time.

Figure 2. Wave-particle duality in atmospheric flows

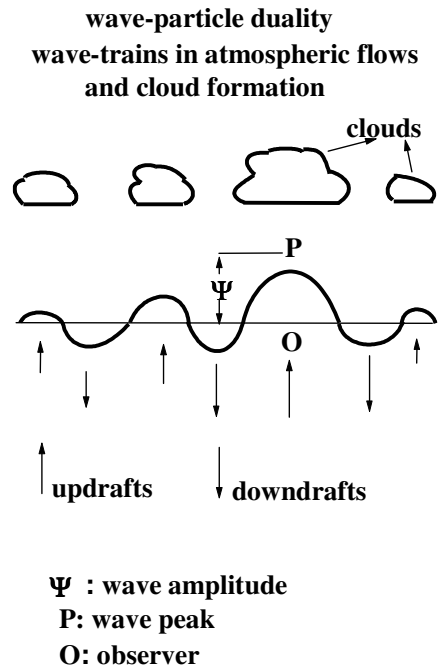

$\Psi$ : wave amplitude
P: wave peak
O: observer

*2.1. Model predictions*

The model predictions [3, 4, and 38] and the interpretation of quantum mechanical laws as applied to macro-scale fluid flows are described in the following. It is shown that the apparent paradoxes of quantum mechanics are physically consistent in the context of atmospheric flows.

(i) Atmospheric flows trace an overall logarithmic spiral trajectory $OR_OR_1R_2R_3R_4R_5$ simultaneously in clockwise and anti-clockwise directions with the quasi-periodic *Penrose tiling pattern* [39] for the internal structure shown in figure 3 (see attached Appendix).

Figure 3. The quasiperiodic *Penrose* tiling pattern

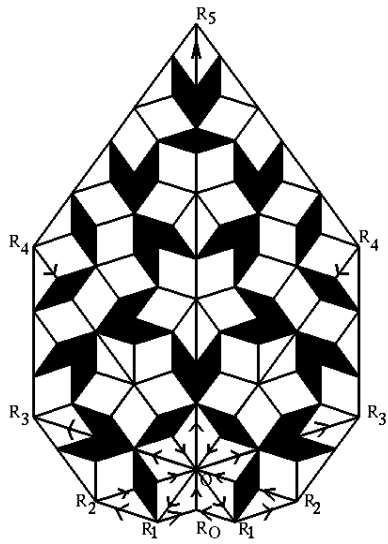

The spiral flow structure can be visualized as an eddy continuum generated by successive length step growths $OR_O$, $OR_1$, $OR_2$, $OR_3$,….respectively equal to $R_1$, $R_2$, $R_3$,….which follow *Fibonacci* mathematical series such that $R_{n+1} = R_n + R_{n-1}$ and $R_{n+1} / R_n = \tau$ where $\tau$ is the *golden mean* equal to $(1+\sqrt{5})/2$ ( $\approx 1.618$). Considering a normalized length step equal to *1* for the last stage of eddy growth, the successively decreasing radial length steps can be expressed as *1, 1/τ, 1/τ², 1/τ³*, ……The normalized eddy continuum comprises of fluctuation length scales *1, 1/τ, 1/τ²*, …….. The probability of occurrence is equal to *1/τ* and *1/τ²* respectively for eddy length scale *1/τ* in any one or both rotational (clockwise and anti-clockwise) directions. Eddy fluctuation length of amplitude *1/τ*, has a probability of occurrence equal to *1/τ²* in both rotational directions, i. e. the square of eddy amplitude represents the probability of occurrence in the eddy continuum. Similar result is observed in the subatomic dynamics of quantum systems which are visualized to consist of the superimposition of eddy fluctuations in wave trains (eddy continuum).

Nonlocal connections are intrinsic to quasiperiodic Penrose titling pattern. The phenomenon known as nonlocality or "action at a distance" characterize quantum systems. Experiments in quantum optics show that two distant events can influence each other instantaneously. Nonlocal connections in quantum systems apparently violate the fundamental theoretical law in modern physics that signal transmission cannot exceed the speed of light. The distinction between locality and nonlocality is related to the concept of a trajectory [40] of a single point object. The instantaneous nonlocal connections in the stringlike energy flow patterns which represent extended objects can be visualized as shown in figure 4.

Figure 4. : Instantaneous non-local connection in atmospheric eddy circulations

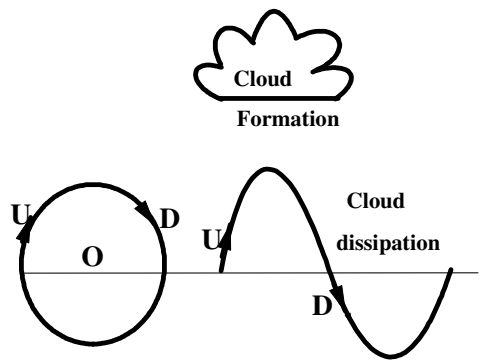

The circulation flow pattern with center *O* and radius *OU* (or *OD*) represents an eddy. In the medium of propagation, namely, atmosphere (air) in this case, upward motion *U* represents convection and cloud formation in association simultaneously with cloud dissipation in downward motion *D*. There is an instantaneous nonlocal connection between the phases of the particles at *U* and *D*. The same concept can be applied to an extended object (figure 2) such as a row of clouds represented by the wave function $\psi$ which represents the superimposition of a continuum of eddies.

In summary, energy pumping at a fundamental frequency generates a broad-band eddy continuum and has been observed as *chaos* in laser and nonlinear optical systems which are basically governed by quantum mechanical laws [41].

(ii) Conventional continuous periodogram power spectral analyses of such spiral trajectories will reveal a continuum of periodicities with progressive increase in phase.

(iii) The broadband power spectrum will have embedded dominant wavebands the bandwidth increasing with period length. The peak periods $E_n$ in the dominant wavebands will be given by the relation

$$E_n = T_s(2+\tau)\tau^n \tag{4}$$

In (4) $\tau$ is the *golden mean* equal to $(1+\sqrt{5})/2$ $[\cong 1.618]$ and $T_s$ is the primary perturbation time period, for example, the solar powered annual cycle (summer to winter) of solar heating in a study of interannual climate variability. Ghil [42] reports that the most striking feature in climate variability on all time scales is the presence of sharp peaks superimposed on a continuous background. The model predicted periodicities are *2.2, 3.6, 5.8, 9.5, 15.3, 24.8, 40.1* and *64.9* years for values of *n* ranging from *-1* to *6*. Periodicities close to model predicted have been reported in weather and climate cycles [43, 44].

Time series analysis of global market economy also exhibits power law behaviour with possible multifractal structure and has suggested an analogy to fluid turbulence. The stock market can be viewed as a self-organizing cooperative system presenting power law distributions, large events in possible co-existence with synchronized behaviour. The observed power law represents structures similar to 'Elliott waves' of technical analysis first introduced in the 1930s. It describes the time series of a stock price as made of different waves; these waves are in relation to each other through the Fibonacci series. The 'Elliott waves' could be a signature of an underlying critical structure of the stock market [45].

(iv) The overall logarithmic spiral flow structure is given by the relation

$$W = \frac{w_*}{k} \ln z \tag{5}$$

In (5) the constant *k* is the steady state fractional volume dilution of large eddy by inherent turbulent eddy fluctuations and *z* is the length scale ratio *R/r*. The constant *k* is equal to $1/\tau^2$ $(\cong 0.382)$ and is identified as the universal constant for deterministic chaos in fluid flows. The steady state emergence of fractal structures is therefore equal to

$$\frac{1}{k} \cong 2.62 \tag{6}$$

The model predicted logarithmic wind profile relationship such as (5) is a long-established (observational) feature of atmospheric flows in the boundary layer, the constant *k*, called the *Von Karman*'s constant has the value equal to *0.38* as determined from observations. Historically, (5), basically an empirical law known as the *universal logarithmic law of the wall*, first proposed in the early 1930s by pioneering aerodynamicists *Theodor von Karman* and *Ludwig Prandtl*, describes shear forces exerted by turbulent flows at boundaries such as wings or fan blades or the interior wall of a pipe. *The law of the wall* has been used for decades by engineers in the design of aircraft, pipelines and other structures [46].

In (5), *W* represents the standard deviation of eddy fluctuations, since *W* is computed as the instantaneous r. m. s. (root mean square) eddy perturbation amplitude with reference to the earlier step of eddy growth. For two successive stages of eddy growth starting from primary perturbation $w_*$, the ratio of the standard deviations $W_{n+1}$ and $W_n$ is given from (5) as *(n+1)/n*. Denoting by $\sigma$ the standard deviation of eddy fluctuations at the reference level *(n=1)* the standard deviations of eddy fluctuations for successive stages of eddy growth are given as integer multiples of $\sigma$, i. e. $\sigma$, $2\sigma$, $3\sigma$, etc. and correspond respectively to *statistical normalised standard deviation*  $t = 0, 1, 2, 3, ....$ (7)

The conventional power spectrum plotted as the variance versus the frequency in log-log scale will now represent the eddy probability density on logarithmic scale versus the standard deviation of the eddy fluctuations on linear scale since the logarithm of the eddy wavelength represents the standard deviation, i. e. the r. m. s. value of eddy fluctuations (5). The r. m. s. value of eddy fluctuations can be represented in terms of statistical normal distribution as follows. A normalized standard deviation $t=0$ corresponds to cumulative percentage probability density equal to *50* for the mean value of the distribution. Since the logarithm of the wavelength represents the r. m. s. value of eddy fluctuations the normalized standard deviation $t$ is defined for the eddy energy as

$$t = \frac{\log L}{\log T_{50}} - 1 \tag{8}$$

In (8) $L$ is the time period (or wavelength) and $T_{50}$ is the period up to which the cumulative percentage contribution to total variance is equal to *50* and $t = 0$. The function $\log T_{50}$ also represents the mean value for the r. m. s. eddy fluctuations and is consistent with the concept of the mean level represented by r. m. s. eddy fluctuations. Spectra of time series of meteorological parameters when plotted as cumulative percentage contribution to total variance versus $t$ have been shown to follow the model predicted universal spectrum [4] which is identified as a signature of quantumlike chaos.

(v) Selvam [47] has shown that (1) represents the universal algorithm for deterministic chaos in dynamical systems and is expressed in terms of the universal *Feigenbaum*'s [14] *constants a* and *d* as follows. The successive length step growths generating the eddy continuum $OR_0R_1R_2R_3R_4R_5$ analogous to the period doubling route to chaos (growth) is initiated and sustained by the turbulent (fine scale) eddy acceleration $w_*$ which then propagates by the inherent property of inertia of the medium of propagation. Therefore, the statistical parameters *mean*, *variance*, *skewness* and *kurtosis* of the perturbation field in the medium of propagation are given by $w_*$, $w_*^2$, $w_*^3$ and $w_*^4$ respectively. The associated dynamics of the perturbation field can be described by the following parameters. The perturbation speed $w_*$ (motion) per second (unit time) sustained by its inertia represents the mass, $w_*^2$ the acceleration or force, $w_*^3$ the angular momentum or potential energy, and $w_*^4$ the spin angular momentum, since an eddy motion has an inherent curvature to its trajectory.

It is shown that *Feigenbaum's* constant $a$ is equal to [47]

$$a = \frac{W_2 R_2}{W_1 R_1} \tag{9}$$

In (9) the subscripts *1* and *2* refer to two successive stages of eddy growth. *Feigenbaum's* constant $a$ as defined above represents the steady state emergence of fractional *Euclidean* structures. Considering dynamical eddy growth processes, *Feigenbaum's* constant $a$ also represents the steady state fractional outward mass dispersion rate and $a^2$ represents the energy flux into the environment generated by the persistent primary perturbation $w_*$ Considering both clockwise and counterclockwise rotations, the total energy flux into the environment is equal to $2a^2$. In statistical terminology, $2a^2$ represents the variance of fractal structures for both clockwise and counterclockwise rotation directions.

The *Feigenbaum's* constant $d$ is shown to be equal to [47]

$$d = \frac{W_2^4 R_2^4}{W_1^4 R_1^4} \tag{10}$$

The equation (10) represents the fractional volume intermittency of occurrence of fractal structures for each length step growth. *Feigenbaum's* constant $d$ also represents the relative spin angular momentum of the growing large eddy structures as explained earlier.

Equation (1) may now be written as

$$2\frac{W^2 R^2}{w_*^2 (dR^2)} = \pi \frac{W^4 R^3}{w_*^4 (dR^3)} \tag{11}$$

In (11) $dR$ equal to $r$ represents the incremental growth in radius for each length step growth, i. e. $r$ relates to the earlier stage of eddy growth.

Substituting the *Feigenbaum's constants a* and *d* defined above in (9) and (10), (11) can be written as

$$2a^2 = \pi d \quad (12)$$

In (12) $\pi d$, the relative volume intermittency of occurrence contributes to the total variance $2a^2$ of fractal structures.

In terms of eddy dynamics, the above equation states that during each length step growth, the energy flux into the environment equal to $2a^2$ contributes to generate relative spin angular momentum equal to $\pi d$ of the growing fractal structures.

It was shown at (6) above that the steady state emergence of fractal structures in fluid flows is equal to $1/k\ (=\tau^2)$ and therefore the *Feigenbaum's constant a* is equal to

$$a = \tau^2 = \frac{1}{k} = 2.62 \quad (13)$$

(vi) The relationship between *Feigenbaum's constant a* and statistical normal distribution for power spectra is derived in the following.

The steady state emergence of fractal structures is equal to the *Feigenbaum's constant a* (6). The relative variance of fractal structure for each length step growth is then equal to $a^2$. The normalized variance $1/a^{2n}$ will now represent the statistical normal probability density for the $n^{th}$ step growth according to model predicted quantumlike mechanics for fluid flows. Model predicted probability density values $P$ are computed as

$$P = \tau^{-4n} \quad (14)$$

or

$$P = \tau^{-4t} \quad (15)$$

In (15) $t$ is the normalized standard deviation (7) and the computed $P$ values are in agreement with statistical normal distribution as shown in Table 1.

**Table 1: Model predicted and statistical normal probability density distributions**

| Growth step | normalized standard dev | probability densities | |
|---|---|---|---|
| | | model predicted | statistical normal |
| $n$ | $t$ | $P = \tau^{-4t}$ | distribution |
| 1 | 1 | .1459 | .1587 |
| 2 | 2 | .0213 | .0228 |
| 3 | 3 | .0031 | .0013 |

The statistical normal distribution characteristics underlie the fractal spacetime continuum fluctuations as shown in the following

The steady state emergence of fractal structures for each length step growth for any one direction of rotation (either clockwise or anticlockwise) is equal to

$$\frac{a}{2} = \frac{\tau^2}{2}$$

since the corresponding value for both direction is equal to $a$ (6).

The emerging fractal spacetime structures have moment coefficient of kurtosis given by the fourth moment equal to

$$\left(\frac{\tau^2}{2}\right)^4 = \frac{\tau^8}{16} = 2.9356 \approx 3$$

The moment coefficient of skewness for the fractal spacetime structures is equal to zero for the symmetric eddy circulations. Moment coefficient of kurtosis equal to *3* and moment coefficient of skewness equal to *zero* characterize the statistical normal distribution underlying the fractal spacetime eddy continuum structure.

Normal distribution characteristics for the eddy continuum fluctuation field can also be derived from model concept as follows.

Let $P$ represent the probability of occurrence in the medium, of bidirectional eddy energy flux with characteristics of a particular large eddy of radius $R$. Since $W$ originates from $w_*$

$$P = \frac{1}{2}\frac{W}{w_*}$$

substituting for $\frac{W}{w_*}$ from (1)

$$P = \frac{1}{\sqrt{2\pi}}\sqrt{\frac{r}{R}}$$

substituting for $\sqrt{\frac{r}{R}}$ from (5)

$$P = \frac{1}{\sqrt{2\pi}}exp\left[-\frac{1}{2}\frac{Wk}{w_*}\right]$$

Substituting for $k$, namely,

$$k = \frac{W}{w_*}\frac{R}{r}$$

The probability $P$ is obtained as

$$P = \frac{1}{2\pi}exp\left[\frac{1}{2}\frac{W^2}{w_*^2}\frac{R}{r}\right] \qquad (16)$$

For any two successive stages of eddy growth (1)

$$W_1^2 R_1 = W_2^2 R_2$$

therefore

$$\frac{W_1^2 R_1^2}{W_2^2 R_2^2} = \frac{R_1}{R_2} = \frac{r}{R}$$

Linearising (16) for two successive stages of eddy growth

$$P = \frac{1}{\sqrt{2\pi}}exp\left[-\frac{1}{2}\frac{R_1}{R_2}\right]$$

$$= \frac{1}{\sqrt{2\pi}}exp\left[-\frac{1}{2}\frac{W_2^2}{W_1^2}\right]$$

Therefore *statistical normal* distribution characteristics [48] are followed by the probability $P$ of occurrence of eddy fluctuation $W$ originating from earlier stage perturbation $w_*$.

(vii) The power spectra of fluctuations in fluid flows can now be quantified in terms of universal *Feigenbaum's constant a* as follows.

The normalized variance and therefore the statistical normal distribution is represented by (14)

$$P = a^{-2t} \qquad (17)$$

In (17) $P$ is the probability density corresponding to normalized standard deviation $t$. The graph of $P$ versus $t$ will represent the power spectrum. The slope $S$ of the power spectrum is equal to

$$S = \frac{dP}{dt} \approx -P \qquad (18)$$

The power spectrum therefore follows inverse power law form, the slope decreasing with increase in $t$. Increase in $t$ corresponds to large eddies (low frequencies) and is consistent with observed decrease in slope at low frequencies in dynamical systems.

(viii) The fractal dimension $D$ can be expressed as a function of the universal *Feigenbaum's constant* $a$ as follows.

The steady state emergence of fractal structures is equal to $a$ for each length step growth (7) and (13) and therefore the fractal structure domain is equal to $a^m$ at $m^{th}$ growth step starting from unit perturbation. Starting from unit perturbation, the fractal object occupies spatial (two dimensional) domain $a^m$ associated with radial extent $\tau^m$ since successive radii follow *Fibonacci* number series. The fractal dimension $D$ is defined as

$$D = \frac{d \ln M}{d \ln R}$$

where $M$ is the mass contained within a distance $R$ from a point in the fractal object. Considering growth from $n^{th}$ to $(n+m)^{th}$ step

$$d \ln M = \frac{dM}{M} = \frac{a^{n+m} - a^n}{a^n} = a^m - 1 \tag{19}$$

similarly

$$d \ln R = \frac{dR}{R} = \frac{\tau^{n+m} - \tau^n}{\tau^n} = \tau^m - 1 \tag{20}$$

Therefore the fractal dimension $D$ is given as

$$D = \frac{\tau^{2m} - 1}{\tau^m - 1} = \tau^m + 1 \tag{21}$$

The fractal dimension increases with the number of growth steps. The dominant wavebands increase in length with successive growth steps. The fractal dimension $D$ indicates the number of periodicities which superimpose to give the observed four dimensional spacetime structure to the flow pattern. The above concept of dimension for real world spacetime patterns is consistent with El Naschie's [15] interpretation of dimensions for superstring theories in particle physics, namely a string rotates in ordinary space and only uses the extra dimensions for vibrations which simulate particle masses. Recent theoretical studies [49] show that the four experienced dimensions (three in space and one in time) are sufficient to describe the observed physical properties of the universe. El Naschie [15] has also derived mathematically and shown that the *golden mean* is intrinsic to the geometry of fractal structures. The correlated fluctuations of the fractal spacetime eddy continuum are analogous to the Bose-Einstein condensation phenomena observed in liquid helium [50 – 53].

(ix) The relationship between *fine structure constant*, i. e. the eddy energy ratio between successive dominant eddies and *Feigenbaum*'s constant $a$ is derived as follows.

$2a^2$ = relative variance of fractal structure (both clockwise and anticlockwise rotation) for each growth step.

For one dominant large eddy (figure 3) $OR_OR_1R_2R_3R_4R_5$ comprising of five growth steps each for clockwise and counterclockwise rotation, the total variance is equal to

$$2a^2 \times 10 \cong 137.07 \tag{22}$$

For each complete cycle (comprising of five growth steps each) in simultaneous clockwise and counterclockwise rotations, the relative energy increase is equal to *137.07* and represents the *fine structure constant* for eddy energy structure. The inverse of this *fine structure constant* will then represent the probability of occurrence of the primary eddy circulation (either clockwise or counter-clockwise) in the dominant large eddy circulation (figure 3) $OR_OR_1R_2R_3R_4R_5$.

Incidentally, the *fine structure constant* in atomic physics [54 – 60], designated as $\alpha^{-1}$, a dimensionless number equal to *137.03604*, is very close to that derived above for atmospheric eddy energy structure. This fundamental constant has attracted much attention [61] and it is felt that quantum mechanics cannot be interpreted properly until such time as we can derive this physical constant from a more basic theory. The laws of physics are in need of a fundamental change [61].

(x) The ratio of proton mass $M$ to electron mass $m_e$, i. e. $M/m_e$ is another fundamental dimensionless number which also awaits derivation from a physically consistent theory. The value of $M/m_e$ determined by observation is equal to about *2000*. In the following it is shown that ratio of energy content of large to small eddies for specific length scale ratios is equivalent to $M/m_e$.

From (22), the energy ratio for two successive dominant eddy growth = $(2a^2 \times 10)^2$

Since each large eddy consists of five growth steps each for clockwise and anticlockwise rotation, the relative energy content of large eddy with respect to primary circulation structure inside this large eddy

$$\frac{(2a^2 \times 10)^2}{10} \cong 1879$$

The above equation also represents the relative upward mass flux (9) for two successive stages of dominant eddy growth with respect to inherent primary turbulent eddy.

The primary circulation corresponds to $OR_OR_1$ (figure 3) with length scale $OR_O$ equal to $\tau^5$ and the dominant large eddy length scale $OR_5$ is then equal to $(\tau^5)^6$. The length scale ratio $OR_5 / OR_O$ is equal to $(\tau^5)^6 / \tau^5 = \tau^{25} \cong 10^{5.22}$. The ratio of the radii of atom and electron is also approximately equal to $10^5$ [62].

Quantum mechanical concepts relating to fundamental particles and universal constants are summarised in the following [57]. The only objects that appear to be exactly the same everywhere are the atoms and their constituent particles. A natural unit of mass is the nucleon mass, equal approximately to that of the hydrogen atom. Nucleons (i. e., protons and neutrons) have a mass *1836* times the mass of the electron.

The constants of nature can be arranged to form natural numbers (often referred to as dimensionless numbers) that are independent of our units of measurement. The ratio of the nucleon and electron masses equal to approximately *1836* is one such number. Another example is the Sommerfeld's fine structure constant defined by $\alpha$

$$\alpha = \frac{2\pi e^2}{hc} \approx \frac{1}{137}$$

In the above equation, *e* is the charge on electron, *h*, the Plank's constant and *c*, the velocity of light. The *fine structure constant* appears whenever radiation interacts with particles, and the combination of *c*, *h* and *e* indicates a wave like (*h*) interaction between particles (*e*) and light (*c*)

The classical electron radius is the size of an electron as calculated prior to the introduction of quantum mechanics. It is obtained by assuming that all the energy $m_e c^2$ of the electron is in the form of electrical energy equal to

$$\frac{e^2}{r}$$

Thus giving a radius *r* expressed by

$$r = \frac{e^2}{m_e c^2} \approx 3 \times 10^{-13} \, cms$$

A characteristic size of atoms is the radius *R* of the hydrogen atom

$$R = \frac{h^2}{m_e e^2} = 0.5 \times 10^{-8} \, cms$$

In the above equation *R* is known as the Bohr orbit radius. The absence of the gravitational constant *G* and *c* indicates that gravity and relativity are not of primary importance in the structure of atoms. The electron radius *r* is given as

$$r = \beta^2 R$$

In the above equation *R* is the radius of the hydrogen atom. Therefore ratio of radii of atom to electron is equal to $\beta^{-2} \approx 10^5$

Summarising [62]

radius of electron ≈ $2.82 \times 10^{-13}$ cms = *r*
radii of most atoms ≈ $2 \times 10^{-8}$ cms = *R*
The scale ratio $z = R/r \approx 10^5$
The radius of the electron is about one hundred-thousandths of the radius of an average atom [62]. The atomic radius is the distance from the atomic nucleus to the outermost stable electron orbital in an electron that is at equilibrium.

The cell dynamical system model concepts therefore enable physically consistent derivation of fundamental constants which define the basic structure of quantum systems. These two fundamental constants could not be derived so far from a basic theory in traditional quantum mechanics for subatomic dynamics [58].

*2.2. No scale model for super gravity*

A no scale energy $M$ can be defined for the large eddy energy with respect to the primary eddy energy $m_p$ [55, 56] as follows

Since phenomenological manifestation of energy occurs only during one half cycle of eddy perturbation, $M$ and $m_p$ can be expressed as equal to the eddy mass fluxes across unit cross-section per second as follows in terms of the respective large and primary eddy r. m. s circulations speeds $W$ and $w_*$.

$$M = \frac{W}{2}$$

and

$$m_p = \frac{w_*}{2}$$

Therefore from (1)

$$M = \frac{1}{2}\sqrt{\frac{2r}{\pi R}}\frac{w_*}{2} = \frac{m_p}{\sqrt{8\pi z}} \qquad (23)$$

The above concept is analogous to the no scale super gravity model of Lahanas and Nanapoulous [63] where $M$, the super-Planck mass is given in terms of the Planck scale $m_p$ ($\approx 10^{19}$ *Gev*) which corresponds to the first excited state of these strings [64]. The virtues of the no scale super gravity model are automatically vanishing cosmological constant (at least at the classical level), dynamical determination of all mass scales in terms of fundamental Planck scale $m_p$ and acceptable low energy phenomenology. The no scale structure is super symmetric since it fuses together the non-trivial internal symmetries of the internal small scale eddies with the spacetime (Poincare) symmetries of the eddy continuum structure and accounts for the observed fractal geometry in nature.

It is possible to compute and show that the Planck length scale $l_p$ is about *20* orders of magnitude smaller than the electron length scale $l_e$ by substituting the known values [62] of $m_p$ equal to *2.176×10⁻⁵* gm and the electron mass $m_e$ equal to *9.109×10⁻²³* gm in (23). Each length step of eddy growth corresponds to angular rotation equal to *36 degrees* as shown in figure 3 for the quasiperiodic Penrose tiling pattern structure traced by the growing large eddy. Starting from unit length scale perturbation, one complete cycle (*360 degrees*) of eddy growth consists of *10* length steps and results in a final length scale equal to $\tau^{10}$, the successive length scales following the Fibonacci number series. Let the primary Planck length scale eddy perturbation of length scale $l_p$ equal to $\tau^{10}$ generate the electron length scale $l_e$ eddy perturbation after $n_e$ number of cyclical growth steps. The length scale ratio $z_{ep}$ of electron with respect to the primary eddy cycle corresponding to Planck length scale is represented by $n_e$ and is equal to about *21* as shown in the following.

$$l_e = \tau^{10n_e}$$

$$z_{ep} \approx \frac{l_e}{l_p} = \frac{\tau^{10n_e}}{\tau^{10}} = \tau^{10(n_e - 1)}$$

$$z_{ep} = \frac{1}{8\pi}\left[\frac{m_p}{m_e}\right]^2 = \frac{1}{8\pi}\left[\frac{2.176 \times 10^{-5}}{9.109 \times 10^{=23}}\right]^2 = \frac{1}{8\pi}\left[23.9 \times 10^{21}\right]^2 = 22.71 \times 10^{42}$$

$$z_{ep} = \tau^{10(n_e - 1)} == 22.71 \times 10^{42}$$

$$log\left[\tau^{10(n_e - 1)}\right] = log\, 22.71 + 42$$

$$10(n_e - 1) \times 0.21 = 1.356 + 42 = 43.356$$

$$2.1 n_e - 2.1 = 43.356$$

$$n_e = \frac{45.456}{2.1} = 21.65 \approx 21$$

Penrose [61] has discussed the mysterious aspects of the observed mass problem in particle physics, namely the macroscopic Planck mass ($2.176 \times 10^{-5}$ gm) is associated with length scale *20* orders of magnitude smaller than the tiniest length scales encountered in particle physics such as that of the electron. The above derivation shows that the Planck distance of *$1.6163 \times 10^{-33}$ cm* [62] is some *20* orders of magnitude smaller than the electron length scale consistent with model concepts (23).

The string theory visualizes particles as extended objects and thereby avoids singularities, a major problem in the application of point-like concept for particles in traditional physics [65].

The string theory for quantumlike mechanics in atmospheric flows is analogues to Bohm's concept of *implicate order* for subatomic dynamics of quantum systems [66].

The cell dynamical system model described above is basically a string theory applicable to all dynamical systems ranging from macro-scale atmospheric flows to subatomic scale quantum systems. The four-dimensional real world spacetime continuum fluctuation are manifestation of the superimposition of a hierarchical continuum of eddy circulations (vortices within vortices), whose centripetal acceleration add cumulatively to represent the inertial mass, which is equivalent to gravitational mass. Recent studies [67] show that in a strong magnetic field, electrons swirl around magnetic field lines, creating a vortex. Under right conditions, a vortex can couple to an electron, acting as a single unit. Vortex geometrical structure is ubiquitous in macro-scale as well as microscopic subatomic dynamical fluctuation patterns. Inside a superconductor, electrical currents flow without resistance. Almost as remarkable as this electron flow without dissipation are the quantized thread-like-vortices of charge that swirl like miniature tornadoes around lines of magnetic field [68].

*2.3. Cantorian fractal spacetime, quantum-like chaos and scale relativity in atmospheric flows*

Cantorian fractal spacetime fluctuations characterize quantumlike chaos in atmospheric flows. The macroscale atmospheric flow structure behaves as a unified whole quantum system, where, the superimposition of a continuum of eddies results in the observed global weather patterns with long-range spatiotemporal correlations such as that of the widely investigated El Nino phenomenon [3]. Large eddies are visualised as envelopes enclosing smaller eddies, thereby generating a hierarchy of eddy circulations, originating initially from a fixed primary small scale energising perturbation, e.g. the frictional upward momentum flux at the boundary layer of the earth's surface. In the following sections it is shown that the relative motion concepts of Einstein's special and general theories of relativity are applicable to eddy circulations originating from a constant primary perturbation.

*2.4. Physical concepts in spacetime relativity*

The equations of motion enunciated by Newton in 1687 [69] were believed to describe nature correctly for over 200 years. The ideas of Newton involve the assumption that the laws of motion, and indeed all the laws of physics, are the same for an observer at "rest" as for an observer moving with uniform velocity with respect to the "rest" system. This symmetry principle is sometimes called the *principle of relativity*. The *principle of relativity* in Newton's and Einstein's theories of mechanics differs only in the way that the speed of the observer affects observations of positions and times in the two theories [26]. If an inertial reference system is defined as one in which Newton's laws describe the behavior of bodies, any other reference system, which moves with constant velocity with respect to this first inertial system, is also an inertial system. Time and space seem to be independent of the particular frame used [70]. The concept of relativity (Galilean), a symmetry principle, has been used in mechanics for a long time. By symmetry is meant an invariance against change, something stays the same in spite of some potentially consequential alteration [71]. Investigations into the phenomenon of electricity and magnetism culminated in 1860 in Maxwell's equations of the electromagnetic field, which describe electricity, magnetism and light in one uniform system. However during the period 1890 - 1905 it was recognized that Maxwell equations did not seem to obey the inherent symmetries present in the laws of motion of Galileo and Newton. One of the consequences of Maxwell's equations is that if there is disturbance in the field such that light is generated, these electromagnetic waves go out in all directions equally at the same speed *c*, equal to about $3 \times 10^5$ km/sec. Another consequence of the equations is that if the source of the disturbance is moving, the light emitted goes through space at the same speed *c*. This is analogous to the case of sound, the speed of sound waves being likewise independent of the motion of the source [72]. Incidentally, the constant *c* happened to be first discovered by workers in the field of electricity, long before electromagnetic waves were known to exist [69].

A number of experiments based on the general idea of Galilean relativity were performed to determine the speed of light. Michelson and Morley, in 1887 found that the velocity of a beam of light moving from east to west is the same as that of a beam of light moving from north to south. The east-west velocity might have been expected to be influenced by the velocity of the earth, but such was not the case. About 20 years later, H. A. Lorentz provided the solution by suggesting that material bodies contract when they are moving and that this foreshortening is only in the direction of the motion and also that if the length is $L_o$ when body is at rest, then when it moves with speed *u* parallel to its length, the new length $L_1$ is given as

$$L_1 = L_0 \sqrt{1 - \frac{u^2}{c^2}}$$

Although the contraction hypothesis successfully accounted for the negative result of the experiment, it was open to the objection that it was invented for the express purpose of explaining away the difficulty and was too artificial [72]. The contraction in length is concomitant with modification in time elapsed by the factor

$$\frac{1}{\sqrt{1 - \frac{u^2}{c^2}}} \qquad (24)$$

i. e., moving clocks run slower.

Based on the above hypothesis of linear contraction and time dilation in moving objects, Lorentz showed that Maxwell's equations retain their symmetry, i. e. remain unchanged when the following Lorentz transformations are applied.

$$x' = \frac{x - ut}{\sqrt{1 - \frac{u^2}{c^2}}}$$

$$y' = y$$

$$z' = z$$

$$t' = \frac{t - \frac{ux}{c^2}}{\sqrt{1 - \frac{u^2}{c^2}}}$$

Lorentz's transformations introduced into the laws of mechanics, the speed of light, basically an electromagnetic constant.

The corresponding Galilean transformations are

$$x' = x - ut$$
$$y' = y$$
$$z' = z$$
$$t' = t$$

These equations relate the space and time coordinates (*x, y, z* and *t*) of a system at rest to those (*x', y', z',* and *t'*) of a system in uniform relative motion of speed *u* in the *x* direction.

Einstein, following a suggestion originally made by Poincare, then proposed in his *special theory of relativity* that all physical laws should be of such a kind that they remain unchanged under a Lorentz transformation [72]. Applying Lorentz transformations to Newton's laws of motion, Einstein, in 1905, showed that the mass *m* in Newton's laws of motion should now be written as

$$m = \frac{m_0}{\sqrt{1 - \frac{u^2}{c^2}}}$$

In the above equation $m_0$ is the rest mass and $c$ is the speed of light equal to about $3 \times 10^5$ km sec$^{-1}$. Einstein's *special theory of relativity* proposed in 1905 introduced modification of laws of motion, related to how physical observers measure spatial displacements and time intervals.

The two basic postulates of Einstein's *special theory of relativity* are as follows [70].
  (i) The laws of electrodynamics and of mechanics are the same in all inertial frames. This includes the requirement that $c$, the velocity of light in free space, is invariant.
  (ii) It is impossible to devise an experiment, which defines a state of absolute motion. There is no special "rest" frame of reference.

*2.5. Special relativity to general relativity*

After proposing the special theory of relativity, Einstein discovered that the inverse square law of gravitation could not co-exist consistently with the *special theory of relativity* [73, 74]. One of the major achievements of the *special theory of relativity* was the demonstration that the speed of light is a constant and imposes a limit to the maximum attainable speed of any material object or electromagnetic wave. This limiting speed $c$ plays a crucial role in relating spacetime measurements of inertial observers, i. e. observers moving under no forces. The gravitational attraction as postulated by Newton clearly exceeded this speed limit, since it was instantaneous. Further, the phenomenon of gravity prevents us from defining the inertial observers in the first place. The inertial observers so fundamental to *special relativity* do not exist because of ever-present force of gravity.

In his *general theory of relativity* proposed in 1915, Einstein gave an ingenious interpretation to this property of gravity. Realizing that gravity is permanently attached to space, he argued that it, in fact describes an intrinsic property of space and time, viz. its geometry. The geometry of space and time must be of a curved non-Euclidean type, i. e. spacetime is curved. By treating spacetime as curved Einstein eliminated gravity as a physical force. In his *general theory of relativity* Einstein gave a set of equations, which relate the geometrical properties of spacetime to the distribution of gravitating matter within it. *Special relativity* brought into physics the important notion that space and time together form a joint entity. The measurements of spatial distances and time intervals in the special theory are performed according to flat space geometry. The notion of curvature of spacetime and its relation to gravity is the remarkable new feature of the *general theory*.

There is only a minute difference between the predictions of general relativity and Newtonian gravity [73, 74].

Nottale [75] has discussed the implications of the fractal spacetime characteristics on fundamental physical laws.

## 3. Scale relativity and fractal spacetime structures in atmospheric flows

Dynamical systems in nature exhibit selfsimilar spacetime fractal fluctuations of all scales down to the microscopic scales of the subatomic world, i. e. the vacuum zero point energy fluctuations. Selfsimilarity implies long-range correlations, i. e. nonlocal connections in space and time. The ubiquitous fractal spacetime structures found in nature imply a self-organization or self-assembly process which is independent of microscopic details such as physical, chemical, physiological, etc of the dynamical system. The cell dynamical system model for atmospheric flows proposed by Selvam and Fadnavis [4, 5] may be directly applicable to all dynamical systems in general and in particular to the subatomic dynamics of quantum systems (Section 2.1). The model is based on the concept [17] that spacetime integration of enclosed small-scale (turbulent) fluctuations results in the formation of large-scale (eddy) circulations (figure 1.).

Large eddies are visualized as envelopes enclosing inherent small scale eddies, thereby generating a continuum of eddies, the spatial integration at each level generating the next level (large scale) and so on. The relationship between the root mean square (r. m. s.) circulation speeds $W$ and $w_*$ respectively of large and small eddies and their respective radii $R$ and $r$ is given as (1)

$$W^2 = \frac{2}{\pi} \frac{r}{R} w_*^2$$

The primary perturbation $w_*$ is constant and generates a continuum of eddies of progressively increasing radii $R$. The r. m. s. circulation speed $W$ at each level represents the mean as well as standard deviation, i. e. it is a relative velocity with respect to the constant generating perturbation $w_*$
Therefore the factor

$$\sqrt{1 - \frac{u^2}{c^2}}$$

in the Lorentz transformation at (24) becomes equal to *1*, since the relative velocity $u = 0$. Also, $w_*$ is equivalent to $c$ and is a constant primary perturbation. The Lorentz transformations when applied to eddy dynamics reduce to classical mechanics of Galileo and Newton, since, by concept the eddy r. m. s. circulation speeds $W$ are relative to the constant primary perturbation speed $w_*$.

Einstein's principles of *special relativity* and *general relativity* are applicable to eddy dynamics as summarized in the model predictions [4, 5] in the following

   (i) Spacetime fractal structures are signatures of string like energy flow in a hierarchy of vortices tracing an overall logarithmic spiral trajectory with the quasiperiodic Penrose tiling pattern for the internal structure.

   (ii) The logarithmic spiral energy flow structure can be resolved as a continuum of eddy circulations, which follow Kepler's laws of planetary motion, in particular the third law. The inertial masses of eddies representing gravitational masses, therefore follow Newton's inverse square law of gravitation. Fractal spacetime fluctuations are related to gravity and is consistent with El Naschie's [15] conjecture that gravity is caused by 'fractal' fluctuations of time.

   (iii) Instantaneous non-local connection, prohibited in Einstein's special theory of relativity, is possible and consistent in the context of eddy circulations which are considered as extended objects as explained in the following. The bidirectional energy flow intrinsic to eddy circulations is associated with bimodal, i. e. formation and dissipation respectively of phenomenological form for manifestation of energy such as the formation of clouds in updrafts associated with simultaneous dissipation of clouds in adjacent downdrafts, thereby generating discrete cellular geometry to cloud structure.

Gravitation is defined as a property of spacetime geometry in Einstein's general theory of relativity. The concept of pointlike particle of zero dimensions in classical physics introduces infinities or singularities in the smooth spacetime fabric representing gravitational field. Further, pointlike particles are associated with trajectories, where, the speed of the particle cannot exceed the speed of light according to Einstein's *special theory of relativity*.

The cell dynamical system model [4, 5] discussed in this section introduces the concept of extended objects thereby avoiding singularities and also possessing instantaneous nonlocal connection.

Cantorian fractal spacetime structure to atmospheric flow patterns is a result of the superposition of a continuum of eddies, which function as a unified whole quantum system. Wave-particle duality in the quantum system of atmospheric flows is a result of bimodal (formation and dissipation) form for manifestation of energy in the bidirectional energy flow intrinsic to eddy circulations, such as the formation of clouds in updrafts and dissipation of clouds in adjacent downdrafts, manifested in the common place occurrence of clouds in a row (figures. 2 and 4).

The eddy circulations follow Kepler's laws of planetary motion, in particular, the third law and therefore Newton's inverse square law for gravitation is applicable to eddy masses. The root mean square (r. m. s) circulation speeds of the eddies are relative to and less than the primary constant perturbation. Therefore, the basic criteria invoked in Einstein's *special and general theories of relativity* are incorporated in the concept for generation of eddy continuum. It is possible that the vacuum zero point electromagnetic field fluctuations, may self-organize to generate particles such as the electrons, protons etc., in a manner similar to the formation of atmospheric eddy continuum. The vacuum may be a permanent nonzero source of energy in the universe [76].

In summary, quantum mechanical laws, which govern the subatomic dynamics of quantum systems such as the photon or electron, are shown to be applicable to macro-scale fluid flow dynamics. The cell dynamical system model for turbulent fluid flows developed by the author shows that quantum mechanical laws are applicable to the energy structure of atmospheric weather systems. Some of the yet unresolved problems of quantum mechanics for the subatomic domain are now shown to be physically consistent commonplace observational facts in the context of atmospheric eddy dynamics as listed in the following: (i) wave-particle duality is inherent to the bi-directional eddy energy circulation and corresponds to the bimodal (formation and dissipation respectively) phenomenological form for energy display, e.g. cloud formation and dissipation respectively in adjacent updraft and downdraft regions in atmospheric flows. Such a concept may possibly explain the essentially wave-like polarization of the particle-like quanta of electromagnetic radiation (ii) spirality or spin is intrinsic to the bi-directional eddy energy flow, the coupled system of updrafts and downdrafts having opposite spins are analogous to the electron-positron pair of the subatomic domain (iii) instantaneous non-local response and adjustment occurs between the spatially separated domains of the low and high pressure centers respectively of the updraft and downdraft regions because of the continuity of each eddy circulation, the downdraft being an instantaneous response to the updraft thereby satisfying Newton's third law of motion. The instantaneous non-local connection (action at a distance) between the two streams of the bi-directional energy flow of each individual eddy circulation is also responsive to eddy circulations in all the other scales because of the co-existence of all the scales of eddies in the unified eddy continuum (overall environment) with implicit ordered energy flow between all the scales. The analogous non-local effects of quantum mechanics for the subatomic phenomena appear as paradoxes and inconsistencies in the absence of the concept of the unified eddy energy structure. The instantaneous non-local connection (or action at a distance) is seen in the Aharonov-Bohm effect and the response of each individual eddy circulation to the overall environment is seen as the Berry's phase. Such non-local connections known as entanglement is a quantum correlation between various parts of a quantum system and is required for processing information [77] (iv) vertical dipole charging of clouds occur in the micro-scale (turbulence) by horizontal convergence field which coincides with the magnetic field of the vertical dipole current, the plane containing the electric and magnetic field being at right angles to the direction of propagation of the large eddy. Therefore the atmospheric large eddy circulations give rise to electromagnetic waves in regions of cloud formation [78]. The magnetic field may therefore be identified with the horizontal convergence (divergence) field generated by the turbulent up (down) drafts. The atmospheric eddy energy structure follows quantum mechanical laws and therefore quantum mechanical and stochastic concepts may be beneficially used in the study of the formation and dynamics of weather systems. Observational results of deterministic chaos in laser emission may be directly applicable to the evolution of atmospheric eddies, i. e. weather patterns [79, 80].

## 4. Conclusions

Meteorologists and climatologists have largely ignored *self-organized criticality* (SOC), a leading candidate for a unified theory of complexity. Theories of complexity, such as SOC, have been underrepresented in the atmospheric sciences because of their "soft science" character. Atmospheric sciences have historically developed from centuries of advancement in the hard sciences, such as physics, mathematics and chemistry, etc. It would have been unlikely to see a quick transition from the classical reductionist and reproducible science approach towards an abstract, holistic and probabilistic complex science [2].

The general systems theory described in this paper for the observed *self-organized criticality* (SOC) is applicable to dynamical systems of all size scales from the microscopic scale subatomic dynamics of quantum systems to macroscale atmospheric flows. The model shows that quantum mechanical laws and gravitational forces are a direct consequence of the fractal structure to space-time continuum fluctuations in dynamical systems of all scales. The idea of using fractal curves in quantum mechanics seems to go back to the work of R. Feynman on path integrals [5].

## Acknowledgement

The author is grateful to Dr. A. S. R. Murty for his keen interest and encouragement during the course of this study.

## Appendix to Section 2.1

### Derivation of the quasicrystalline structure of the quasiperiodic Penrose tiling pattern

The turbulent eddy circulation speed and radius increase with the progressive growth of the large eddy [3]. The successively larger turbulent fluctuations, which form the internal structure of the growing large eddy, may be computed (1) as

$$w_*^2 = \frac{\pi}{2} \frac{R}{dR} W^2 \qquad (25)$$

During each length step growth *dR*, the small-scale energizing perturbation $W_n$ at the $n^{th}$ instant generates the large-scale perturbation $W_{n+1}$ of radius *R* where $R = \sum_{1}^{n} dR$ since successive length-scale doubling gives rise to *R*. Equation (25) may be written in terms of the successive turbulent circulation speeds $W_n$ and $W_{n+1}$ as

$$W_{n+1}^2 = \frac{\pi}{2} \frac{R}{dR} W_n^2 \qquad (26)$$

The angular turning $d\theta$ inherent to eddy circulation for each length step growth is equal to *dR/R*. The perturbation *dR* is generated by the small-scale acceleration $W_n$ at any instant *n* and therefore $dR = W_n$. Starting with the unit value for *dR* the successive $W_n$, $W_{n+1}$, *R*, and $d\theta$ values are computed from (26) and are given in Table 2.

Table 2. The computed spatial growth of the strange-attractor design traced by the macro-scale dynamical system of atmospheric flows as shown in figure 3.

| *R* | $W_n$ | *dR* | $d\theta$ | $W_{n+1}$ | $\theta$ |
|---|---|---|---|---|---|
| 1 | 1 | 1 | 1 | 1.254 | 1 |
| 2 | 1.254 | 1.254 | 0.627 | 1.985 | 1.627 |
| 3.254 | 1.985 | 1.985 | 0.610 | 3.186 | 2.237 |
| 5.239 | 3.186 | 3.186 | 0.608 | 5.121 | 2.845 |
| 8.425 | 5.121 | 5.121 | 0.608 | 8.234 | 3.453 |
| 13.546 | 8.234 | 8.234 | 0.608 | 13.239 | 4.061 |
| 21.780 | 13.239 | 13.239 | 0.608 | 21.286 | 4.669 |
| 35.019 | 21.286 | 21.286 | 0.608 | 34.225 | 5.277 |
| 56.305 | 34.225 | 34.225 | 0.608 | 55.029 | 5.885 |
| 90.530 | 55.029 | 55.029 | 0.608 | 88.479 | 6.493 |

It is seen that the successive values of the circulation speed *W* and radius *R* of the growing turbulent eddy follow the Fibonacci mathematical number series such that $R_{n+1} = R_n + R_{n-1}$ and $R_{n+1}/R_n$ is equal to the golden mean $\tau$, which is equal to $[(1 + \sqrt{5})/2] \cong (1.618)$. Further, the successive *W* and *R* values form the geometrical progression $R_O (1 + \tau + \tau^2 + \tau^3 + \tau^4 + ....)$ where $R_O$ is the initial value of the turbulent eddy radius.

Turbulent eddy growth from primary perturbation $OR_O$ starting from the origin O (figure 3) gives rise to compensating return circulations $OR_1R_2$ on either side of $OR_O$, thereby generating the large eddy radius $OR_1$ such that $OR_1/OR_O = \tau$ and $R_OOR_1 = \pi/5 = R_OR_1O$. Therefore, short-range circulation balance requirements generate successively larger circulation patterns with precise geometry that is governed by the *Fibonacci* mathematical number series, which is identified as a signature of the universal period doubling route to chaos in fluid flows, in particular atmospheric flows. It is seen from figure 3 that five such successive length step growths give successively increasing radii $OR_1$, $OR_2$, $OR_3$, $OR_4$ and $OR_5$ tracing out one complete vortex-roll circulation such that the scale ratio $OR_5/OR_O$ is equal to $\tau^5 = 11.1$. The envelope $R_1R_2R_3R_4R_5$ (figure 3) of a dominant large eddy (or vortex roll) is found to fit the logarithmic spiral $R = R_O e^{b\theta}$ where $R_O = OR_O$, $b = \tan \alpha$ with $\alpha$ the crossing angle equal to $\pi/5$, and the angular turning $\theta$ for each length step growth is equal to $\pi/5$. The successively larger eddy radii may be subdivided again in the golden mean ratio. The internal structure of large-eddy circulations is, therefore, made up of balanced small-scale circulations tracing out the well-known quasi-periodic *Penrose* tiling pattern identified as the quasi-crystalline structure in condensed matter physics. A complete description of the atmospheric flow field is given by the quasi-periodic cycles with *Fibonacci* winding numbers.